\documentclass[twocolumn]{aastex63}

\journalinfo{The Astrophysical Journal Letters}

\usepackage{amssymb, amsfonts, amsmath,framed}

\usepackage{bm}
\expandafter\ifx\csname package@font\endcsname\relax\else
 \expandafter\expandafter
 \expandafter\usepackage
 \expandafter\expandafter
 \expandafter{\csname package@font\endcsname}
\fi
\expandafter\ifx\csname package@font\endcsname\relax\else
 \expandafter\expandafter
 \expandafter\usepackage
 \expandafter\expandafter
 \expandafter{\csname package@font\endcsname}
\fi
\def\bi{\begin{itemize}}
\def\ei{\end{itemize}}
\def\bq{\begin{equation}}
\def\eq{\end{equation}}
\def\bqy{\begin{eqnarray}}
\def\eqy{\end{eqnarray}}

\begin{document}
\title{\large{\textbf{Ion and Electron Acceleration in Fully Kinetic Plasma Turbulence}}}

\correspondingauthor{}
\email{luca.comisso@columbia.edu}

\author{Luca Comisso}
\affiliation{Department of Astronomy and Columbia Astrophysics Laboratory, Columbia University, New York, NY 10027, USA}

\author{Lorenzo Sironi}
\affiliation{Department of Astronomy and Columbia Astrophysics Laboratory, Columbia University, New York, NY 10027, USA}

\begin{abstract}
Turbulence is often invoked to explain the origin of nonthermal particles in space and astrophysical plasmas.
By means of 3D fully kinetic particle-in-cell simulations, we demonstrate that turbulence in  low-$\beta$ plasmas ($\beta$ is the ratio of plasma pressure to magnetic pressure) accelerates ions and electrons into a nonthermal energy distribution with a power-law energy range. The ion spectrum is harder than the electron one, and both distributions get steeper for higher $\beta$.
We show that the energization of electrons is accompanied by a significant energy-dependent pitch-angle anisotropy, with most electrons moving parallel to the local magnetic field, while ions stay roughly isotropic. We demonstrate that particle injection from the thermal pool occurs in regions of high current density. Parallel electric fields associated with magnetic reconnection are responsible for the initial energy gain of electrons, whereas perpendicular electric fields control the overall energization of ions. 
Our findings have important implications for the origin of nonthermal particles in space and astrophysical plasmas.

$\,$

\vspace{0.75cm}

\end{abstract}

\section{Introduction}
Nonthermal energetic particles are ubiquitous in space and astrophysical environments. The solar corona and wind \citep{McComas19}, supernova remnants \citep{reynolds08}, accretion disk coronae \citep{YN14}, jets from supermassive black holes \citep{blandford2019},  and galaxy clusters \citep{brunetti14} are just a few examples where the presence of energetic particles, often in the form of power-law distributions, is observed or inferred. 
However, their origin is still poorly understood. Among the potential processes responsible for their acceleration, turbulence in collisionless plasmas is often invoked as a prime candidate \citep{melrose80,lazarian12,Petrosian12}.

The idea that magnetized turbulence could lead to a power-law energy distribution of accelerated particles can be traced back to Fermi's original model of stochastic particle acceleration through repeated interactions with moving magnetized structures  \citep{fermi49}.  Following Fermi's insight, the development of quasilinear theory \citep[e.g.][]{kennel66,hall67,kulsrud71,achterberg81,jaekel92} and its nonlinear extensions \citep[e.g.][]{volk73,BM97,matthaeus03,shalchi06,YL08} have deepened our understanding of particle acceleration in turbulence. On the other hand, these simplified analytical models neglect some important aspects of the physics of  turbulence. Above all, they ignore the role of magnetic reconnection, which is an essential component of particle acceleration in a magnetized turbulent cascade \citep{arzner04,dmitruk04,Kowal12,ComissoSironi18,ComissoSironi19}.

In order to address this problem in all its complexity, fully kinetic numerical simulations are necessary.  
This has recently been possible in the relativistic  regime \citep{zhdankin17,ComissoSironi18,ComissoSironi19,Zhda2018,zhdankin19,Comisso20,Wong2020,ComissoSironi21,NB2021,Sobacchi2021a}, where the Alfv{\'e}n speed approaches the speed of light and computational costs are substantially reduced.
In the non-relativistic regime, progress has been made through test-particle simulations in synthetic turbulence \citep[e.g.][]{micha96,arzner06,sullivan09,teraki19} or fluid-type simulations \citep[e.g.][]{arzner04,dmitruk04,lehe09,Kowal12,dalena14,lynn14,beresnyak16,gonzalez17,isliker17,Kimura19,Trotta20,Sun21,Pezzi22}, and, more recently, through hybrid fluid-kinetic simulations \citep[e.g.][]{kunz16,Pecora18,arza19}. 
However, these approaches suffer from important limitations: test-particle calculations neglect the back-reaction of accelerated particles onto the electromagnetic fields, and they need to employ ad hoc prescriptions for  particle injection; the hybrid fluid-kinetic approach describes electrons merely as a fluid. The kinetic physics of both ions and electrons is needed to properly account for magnetic reconnection in a collisionless plasma and to self-consistently capture the energization of both species.

In this Letter, we investigate the self-consistent acceleration of  ions and electrons in a turbulent plasma. We consider a low-$\beta$ plasma, which is relevant for various space and astrophysical systems, such as the solar corona or accretion disk coronae. We demonstrate that turbulence naturally produces nonthermal tails with a power-law energy range for both ions and electrons. We show that particles are extracted from the thermal pool in regions of high current density associated with magnetic reconnection, and that the different acceleration mechanisms responsible for the energization of ions and electrons leave a distinctive signature on their pitch-angle distributions.

\section{Numerical Method and Setup}
We adopt a first-principles approach by solving the coupled Vlasov-Maxwell system of equations through the particle-in-cell (PIC) method \citep{birdsall85} with TRISTAN-MP \citep{buneman93,spitkovsky05}. 
We perform the numerical simulations in a triply periodic cubic domain $L^3$ that is discretized into a regular lattice of $1400^3$ cells. 
The initial conditions correspond to uniform plasma with total particle density $2n_0$, Maxwellian-distributed ions and electrons of equal temperature $T_{i0}=T_{e0}=T_0$, and uniform guide magnetic field ${\bm{B}}_0=B_0{\bm{\hat z}}$. 
Turbulence is seeded by initializing a spectrum of magnetic fluctuations having polarizations transverse to ${\bm{B}}_0$ (see \citet{ComissoSironi18,ComissoSironi19} for details). The root-mean-square amplitude of the fluctuations is $\delta B_{0} = B_0$, where $\delta B_{0} = \langle {\delta {B^2} (t=0)} \rangle^{1/2}$. The energy-carrying scale is $l = L/3$.

We resolve the electron skin depth $d_{e}=c/\omega_{pe}$ with $3.3$ cells and employ an average of $64$ computational particles per cell.  
In order to capture the full turbulent cascade from MHD scales to electron kinetic scales, we use a reduced ion-to-electron mass ratio of $m_i/m_e=50$, which gives a domain size $L/d_i=60$, where $d_i=c/\omega_{pi}$ is the ion inertial length, with $\omega_{pi} = ({4\pi n_0 {e^2}/{m_i}})^{1/2}$ indicating the ion plasma frequency. The  ratio of electron plasma frequency $\omega_{pe} =({4\pi n_0 {e^2}/{m_e}})^{1/2}$ to electron gyrofrequency $\omega_{ce} = e B_0/m_e c$ is $\omega_{pe}/\omega_{ce} = 1$, as expected in the solar corona. 
Then, the Alfv{\'e}n speed is $v_A = B_0/({4 \pi n_0 m_i})^{1/2} = 0.14c$.
We consider a low-$\beta$ plasma, with $\beta_0$ ranging from $0.32$ down to $0.04$. Here, $\beta_0 = \beta_{i0}+\beta_{e0}$ is the initial total plasma $\beta$, with $\beta_{i0} = \beta_{e0} = 8 \pi n_0 k_B T_0/B_0^2$. We scan $\beta_0$ by changing the initial plasma temperature $T_0$, which is taken to be $k_B T_0 = (0.08, 0.04, 0.02, 0.01) {m_e c^2}$, where $k_B$ indicates the Boltzmann constant. The resulting $\beta_0$ values of our simulations are $\beta_0 \in \left\{ {0.32, 0.16, 0.08, 0.04} \right\}$. We take the simulation with $\beta_0=0.08$ as the fiducial simulation.

\section{Results}

\begin{figure}
\begin{center}
\includegraphics[width=8.65cm]{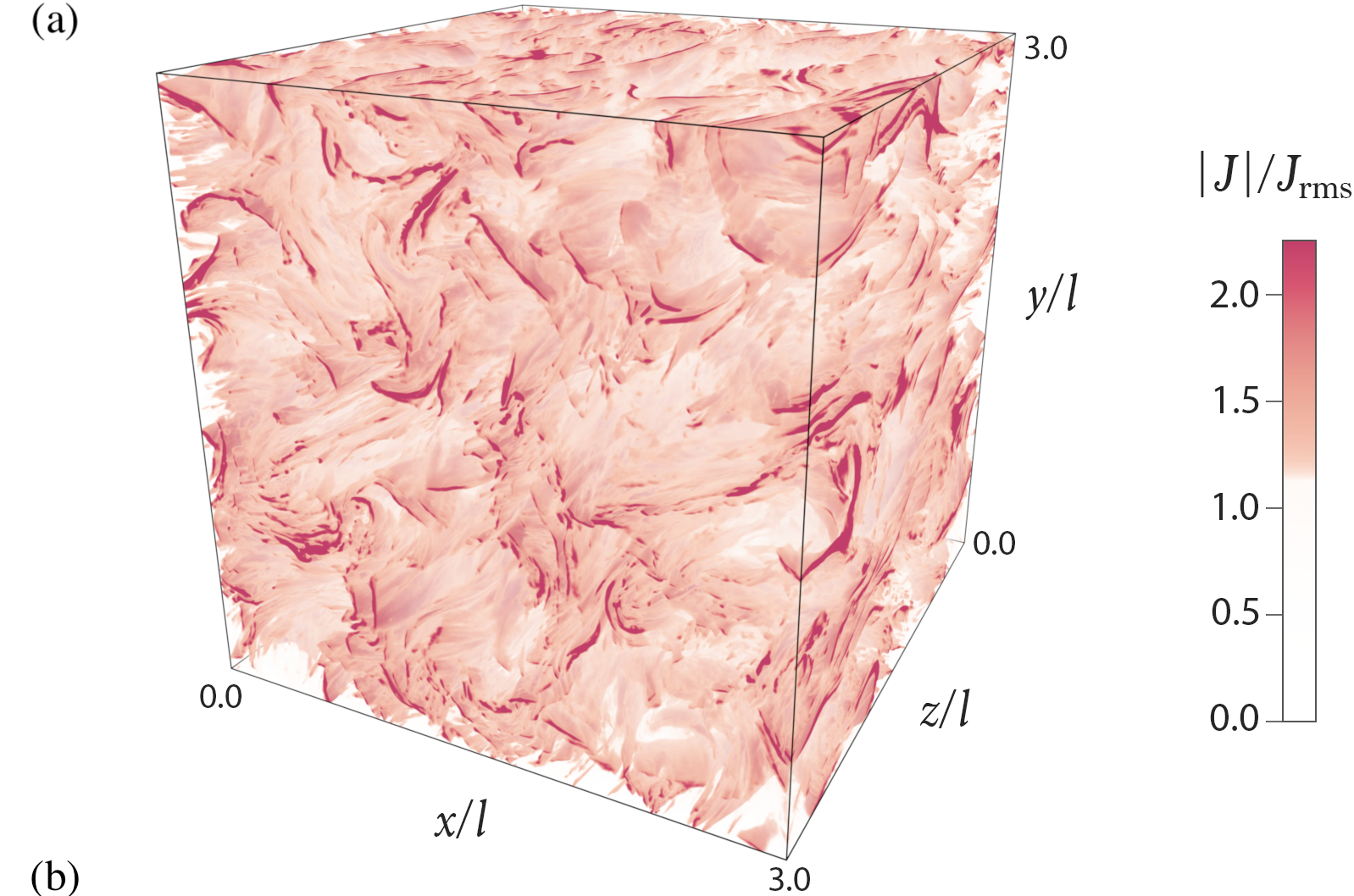}
\includegraphics[width=8.65cm]{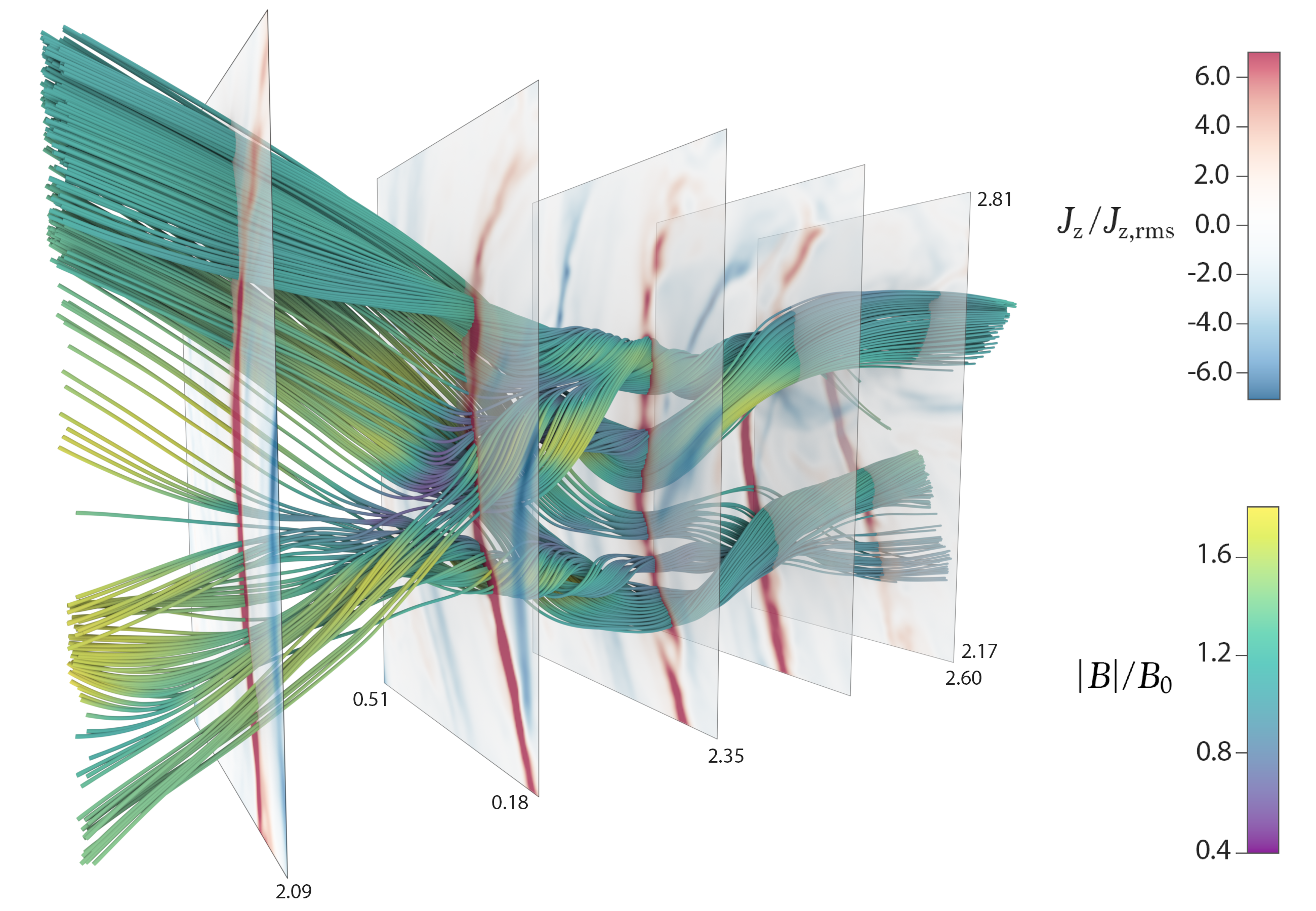}

\vspace{0.415cm}\includegraphics[width=8.65cm]{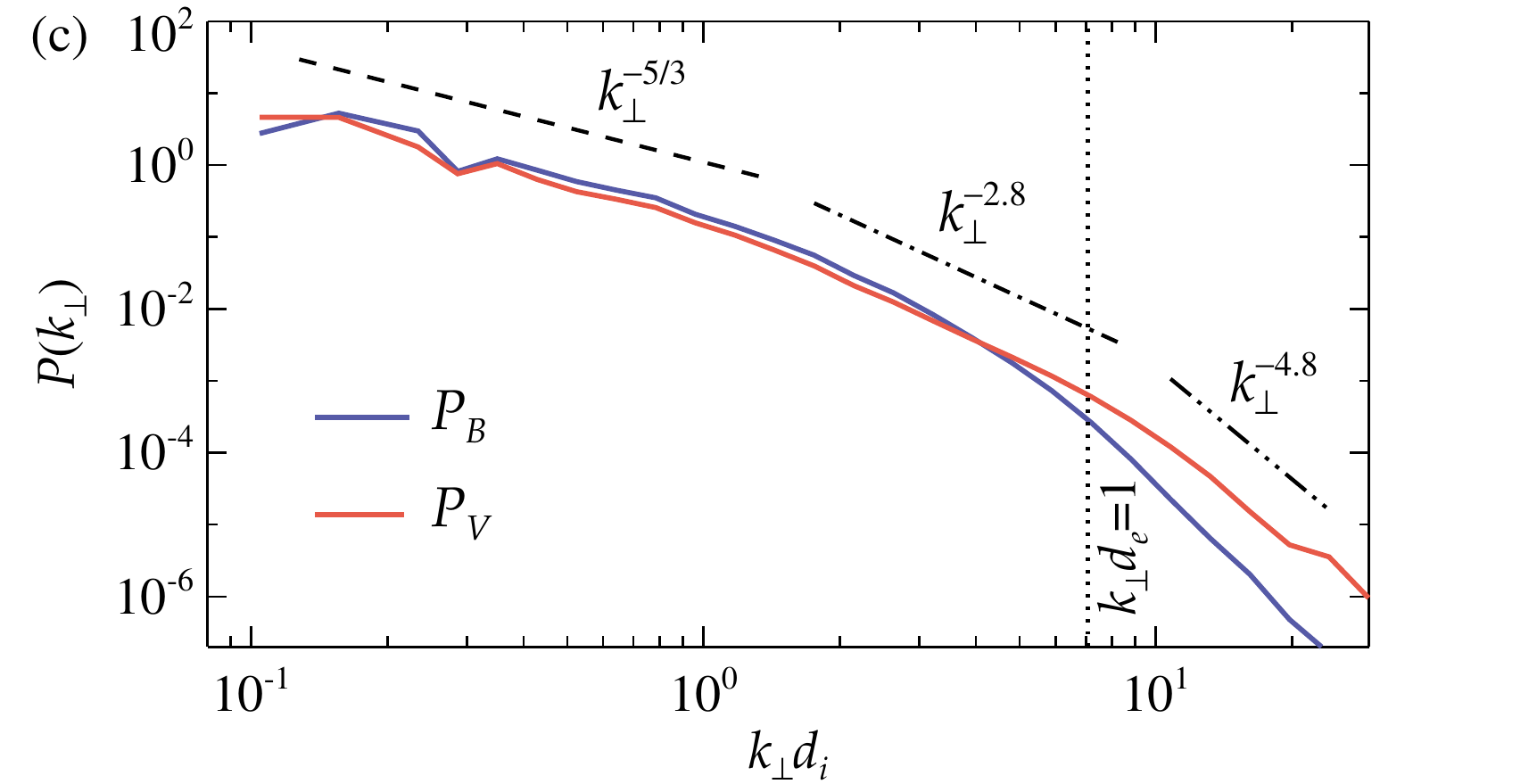}
\end{center}
\vspace{-0.3cm}
\caption{(a) Volume rendering of the current density $|{\bm{J}}|$ at $t = 1.25 \, l/v_A$ from the fiducial simulation ($\beta_0=0.08$).  
(b) Zoomed-in subdomain with five $x$-$y$ slices at different $z$ showing the current density $J_z$, along with selected magnetic field lines illustrating the presence of magnetic flux ropes. 
(c) One-dimensional $k_\bot$ energy spectra of magnetic (blue) and velocity (red) fluctuations at $t = 1.25 \, l/v_A$. Different spectral slopes are provided for reference.}
\label{fig1}
\end{figure}

In Fig.~\ref{fig1}(a), we show a volume rendering of the current density  (normalized to $J_{\rm{rms}} = \langle {J^2} \rangle^{1/2}$) from the fiducial simulation. Sheet-like current density structures are ubiquitous in the turbulent domain. Due to the presence of the mean field ${\bm{B}}_0  = B_0 {\bm{\hat z}}$, these structures are mostly elongated along ${\bm{\hat z}}$. As a result of magnetic reconnection, long and thin current sheets are characterized by the appearance of flux ropes, as shown in Fig.~\ref{fig1}(b). As we show below, magnetic-field-aligned electric fields ($E_\parallel$) associated with magnetic reconnection play an important role in the early stages of electron energization. Finally, in Fig.~\ref{fig1}(c), we show the one-dimensional spectra of the magnetic ($\delta{\bm{B}}$) and bulk velocity ($\delta{\bm{V}}$) fluctuations as a function of $k_\bot = {(k_x^2 + k_y^2)^{1/2}}$. Both power spectra approximately follow $P_{V,B}(k_\bot) \propto k_\bot^{-5/3}$ \citep{GS95} at $k_\bot d_{i} \lesssim 1$. At sub-$d_{i}$ scales, $P_V(k_\bot) \propto k_\bot^{-2.8}$ down to $k_\bot d_{e} \sim 1$  \citep[e.g.][]{Alex09,Chen14}, and it further steepens for $k_\bot d_{e} \gtrsim 1$. 
A similar trend characterizes the $P_{B}(k_\bot)$ spectrum, with the main difference that the steeper range occurs already at $k_\bot d_{e} \gtrsim 0.3$.

\begin{figure}
\begin{center}
\includegraphics[width=8.65cm]{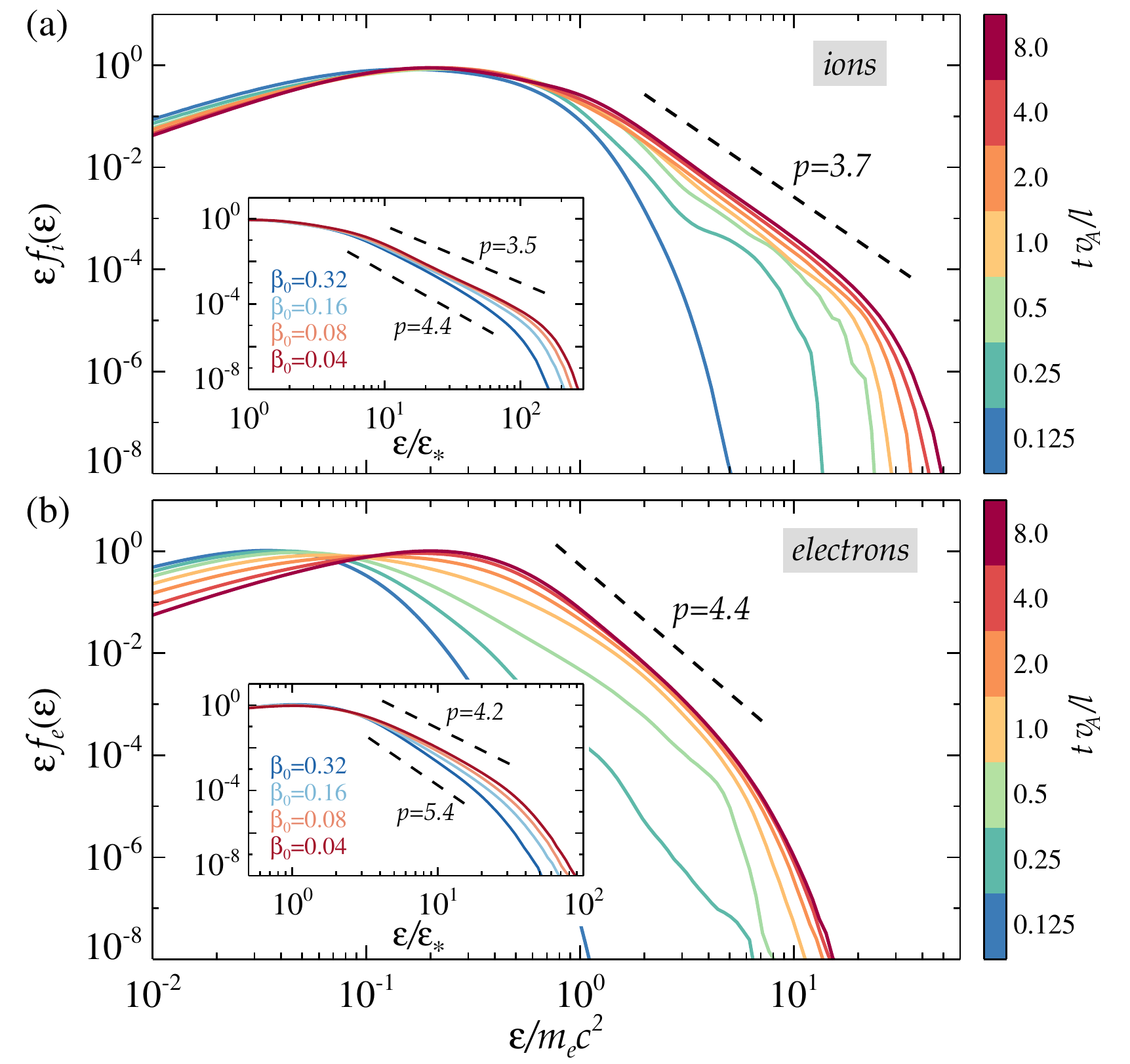}
\end{center}
\vspace{-0.3cm}
\caption{Evolution of the particle energy spectra for (a) ions ($s=i$) and (b) electrons ($s=e$) from the fiducial simulation ($\beta_0=0.08$). Power-law slopes $p=-d\log f_s/d\log \varepsilon$ for the nonthermal tail at late times are shown for reference.
The insets show the late-time ion (top) and electron (bottom) energy spectra for simulations with $\beta_0 \in \left\{ {0.32, 0.16, 0.08, 0.04} \right\}$.}
\label{fig2}
\end{figure}

We show that a key outcome of the turbulent  cascade is the generation of nonthermal particles for both ions and electrons. 
This is illustrated in Fig.~\ref{fig2}, where we show the time evolution of the ion and electron energy spectra $f_s (\varepsilon) = dN(\varepsilon)/d\varepsilon$ (compensated by $\varepsilon$ to emphasize the particle content) for the fiducial simulation. Here, $\varepsilon = (\gamma - 1) m_s c^2$ is the particle kinetic energy ($s=i,e$ for ions or electrons), with $\gamma=(1-v^2/c^2)^{-1/2}$ indicating the particle Lorentz factor.
As a result of turbulent field dissipation, the spectral peak shifts to higher energies. The shift is bounded by the magnetic energy available per each ion-electron pair, $\Delta \overline{\varepsilon}_{i,e} = {\delta B}_0^{\,2}/8\pi n_0 = ({\delta B}_0/B_0)^2 k_B T_{i0}/\beta_{i0} = 0.5 \, m_e c^2$. In addition, a significant number of particles increase in energy by $\Delta \varepsilon \gg \Delta \overline{\varepsilon}_{i,e}$. At late times, when most of the turbulent energy has decayed, the spectra of both ions and electrons extend well beyond the peak into a nonthermal tail of high-energy particles that can be described by a power law $f_s (\varepsilon) d\varepsilon \propto \varepsilon^{-p} d\varepsilon$ for $\varepsilon_{\rm min}  <  \varepsilon  <  \varepsilon_{\rm max}$, 
and a cutoff for $\varepsilon \geq \varepsilon_{\rm max}$.

\begin{figure}
\begin{center}
\includegraphics[width=8.65cm]{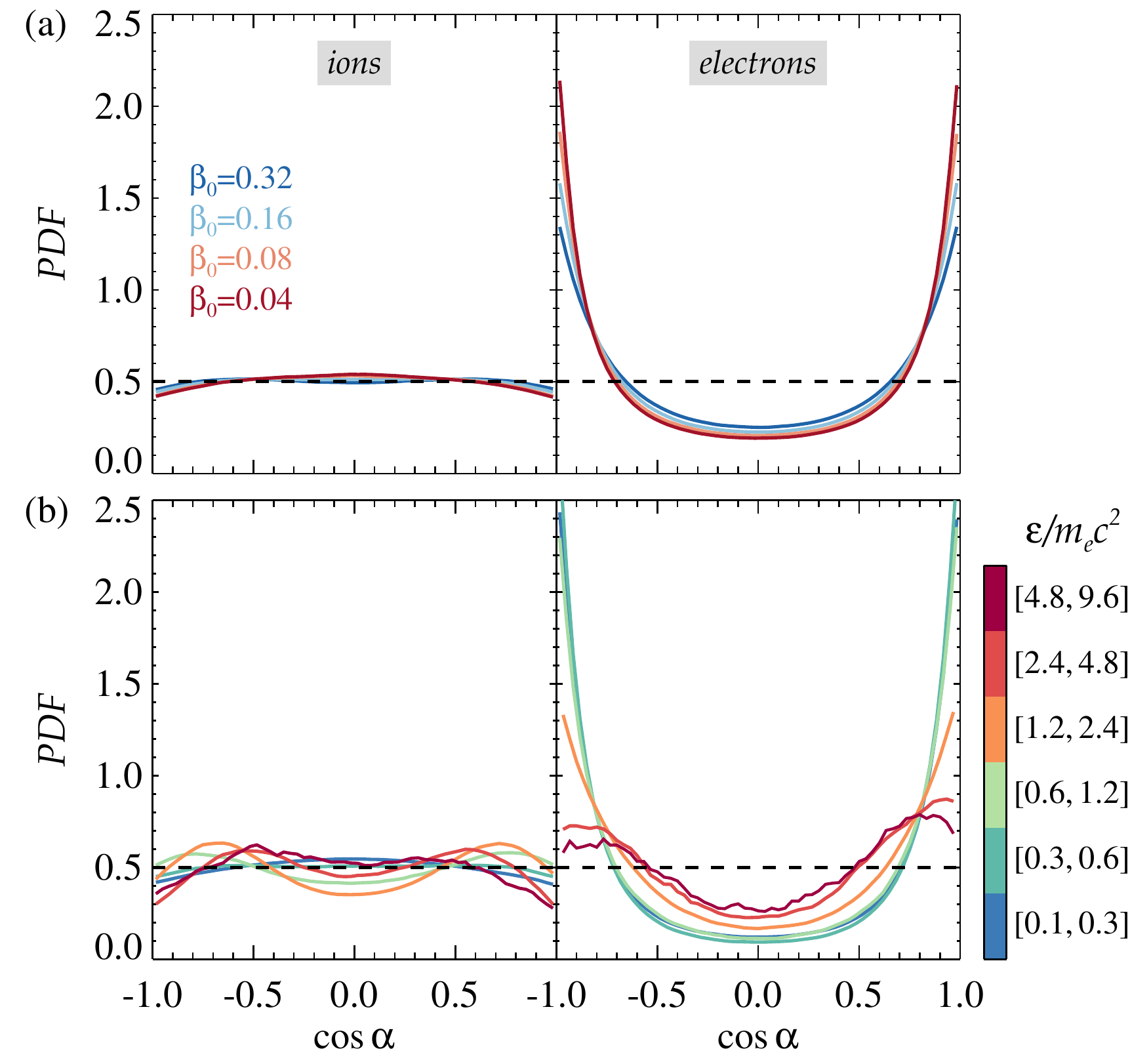}
\includegraphics[width=8.65cm]{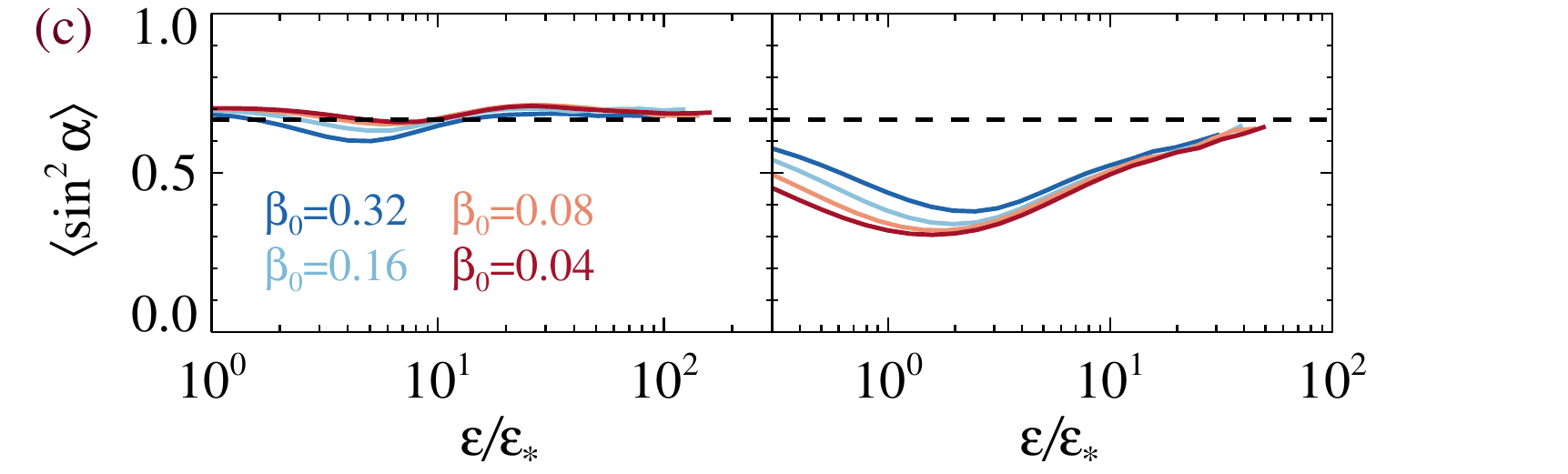}
\end{center}
\vspace{-0.3cm}
\caption{(a) PDFs of $\cos \alpha$ for simulations with $\beta_0 \in \left\{ {0.32, 0.16, 0.08, 0.04} \right\}$. (b) PDFs of $\cos \alpha$ for particles with kinetic energy in different energy intervals for the simulation with $\beta_0=0.08$. (c) $\langle \sin^2 \alpha \rangle$ as a function of particle kinetic energy (normalized by $\varepsilon_*$) for simulations with different $\beta_0$. All plots are obtained from data in the time range 
$t v_A/l \in [2,6]$. Horizontal dashed black lines indicate the expectation for isotropic particles.}
\label{fig3}
\end{figure}

The slope $p$ of the power law depends on $\beta_0$. This is shown in the insets (top for ions, bottom for electrons) of Fig.~\ref{fig2}. 
To facilitate the comparison of the different $\beta_0$ cases, we normalize $\varepsilon$ by the mean energy per particle after turbulent dissipation, $\varepsilon_* = {\varepsilon}_{{\rm th},0} + \kappa \, \Delta \overline{\varepsilon}_{i,e} = (3/2 + 2 \kappa/\beta_0)k_B T_0$, where ${\varepsilon}_{{\rm th},0} = (3/2)k_B T_{0}$ is the initial thermal energy and $\kappa$ is the fraction of magnetic energy transferred to a given species (we take $\kappa = 1/3$ for both ions and electrons, since about $2/3$ of the turbulent magnetic energy is equally distributed between ions and electrons by the end of the simulations). As observed in simulations of low-$\beta$ magnetic reconnection \citep{Dahlin15,XLi19,Zhang21}, the spectrum hardens with decreasing $\beta_0$, with the ion slope approaching $p \approx 3.5$ and the electron slope approaching $p \approx 4.2$ for $\beta_0 = 0.04$.   
 \footnote{The power-law slope depends also on ${\delta B}_0/B_0$, and one might speculate that previous PIC and hybrid-PIC simulations of plasma turbulence \citep[e.g.][]{Groselj18,arza19,Cerri21} have not identified the development of nonthermal power-law tails due to the lower ${\delta B}_0/B_0$ values considered in those simulations, leading to steeper slopes (see \citet{ComissoSironi18,ComissoSironi19,Comisso20} for the effects of ${\delta B}_0/B_0$ on the particle energy distribution in magnetically dominated plasmas).}

An important difference between energized ions and electrons is encoded by the distribution of the pitch angle $\alpha$, i.e., the angle between the particle velocity and the local magnetic field. In Fig.~\ref{fig3}(a), we show the probability density functions (PDFs) of the pitch-angle cosine $\cos \alpha  = {\bm{v}} \cdot {\bm{B}} /({\left| {\bm{v}} \right|\left| {\bm{B}} \right|})$. While ions exhibit roughly isotropic PDFs, electrons are highly anisotropic with PDFs strongly peaked at $\cos \alpha = \pm 1$, i.e., electrons move mostly along the magnetic field lines. 
The different $\beta_0$ cases display a similar behavior, apart from the fact that electrons have higher PDF peaks near $\cos \alpha = \pm 1$ for lower $\beta_0$. As we show below, this trend is due to the fact that electrons gain more energy from $E_\parallel$ as $\beta_0$ decreases.

The PDFs shown in Fig.~\ref{fig3}(a) are dominated by particles near the peak of $\varepsilon f_s (\varepsilon)$, since these particles control the number census. To characterize the anisotropy of higher-energy particles, we construct PDFs of $\cos \alpha$ for particles in different energy ranges. Fig.~\ref{fig3}(b) shows those PDFs for the fiducial simulation. At moderate energies, $\varepsilon \lesssim {\varepsilon}_{{\rm th},0} + \Delta \overline{\varepsilon}_{i,e}$, ions and electrons display PDFs that are similar to those in Fig.~\ref{fig3}(a). However, at higher energies the PDFs peak at intermediate values between $\cos \alpha = 0$ and $\cos \alpha = \pm 1$. This is particularly prominent for electrons, which lose the ${\bm{B}}$-field alignment that characterizes low and moderate energies.

\begin{figure}[]
\begin{center}
\includegraphics[width=8.65cm]{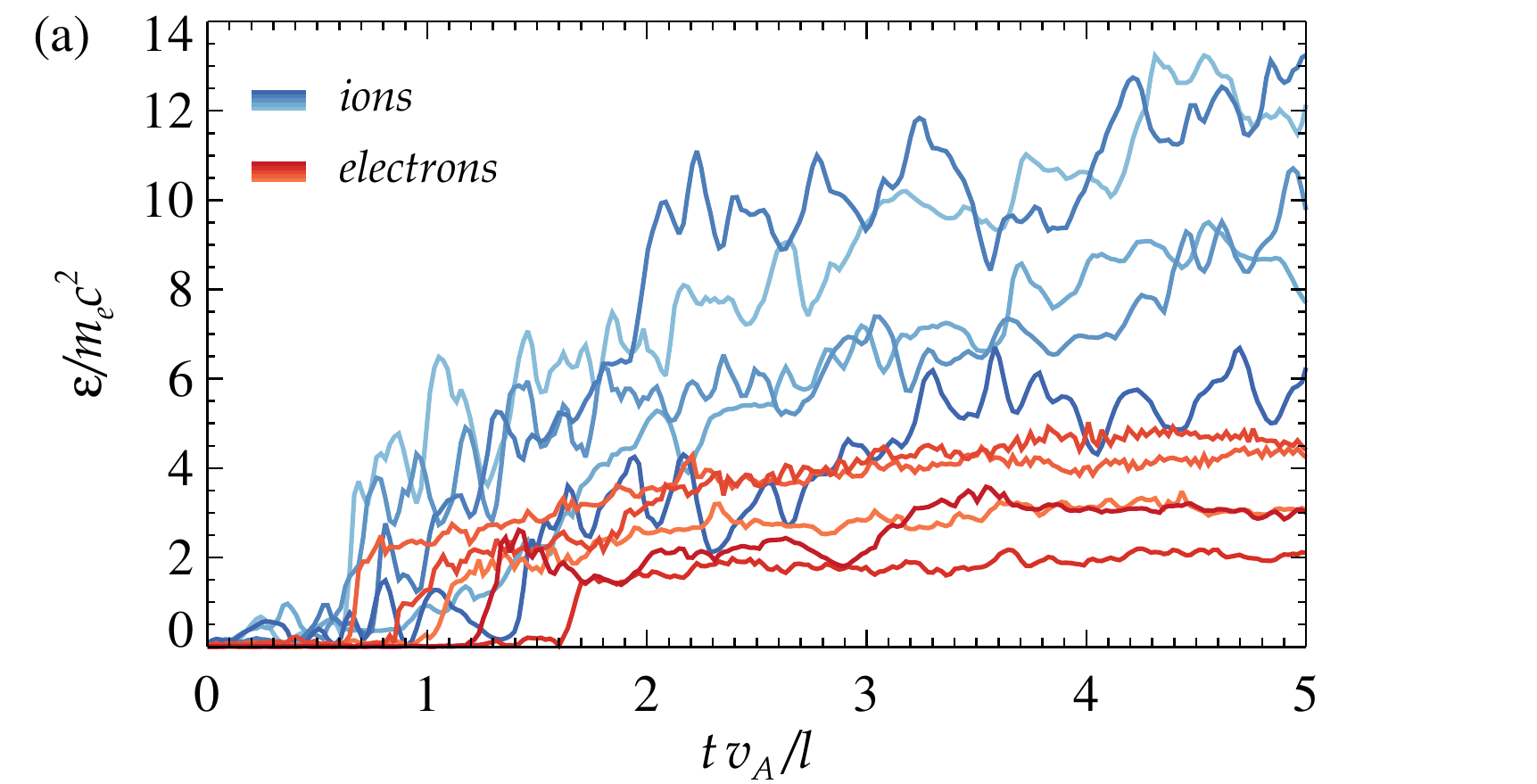}
\includegraphics[width=8.65cm]{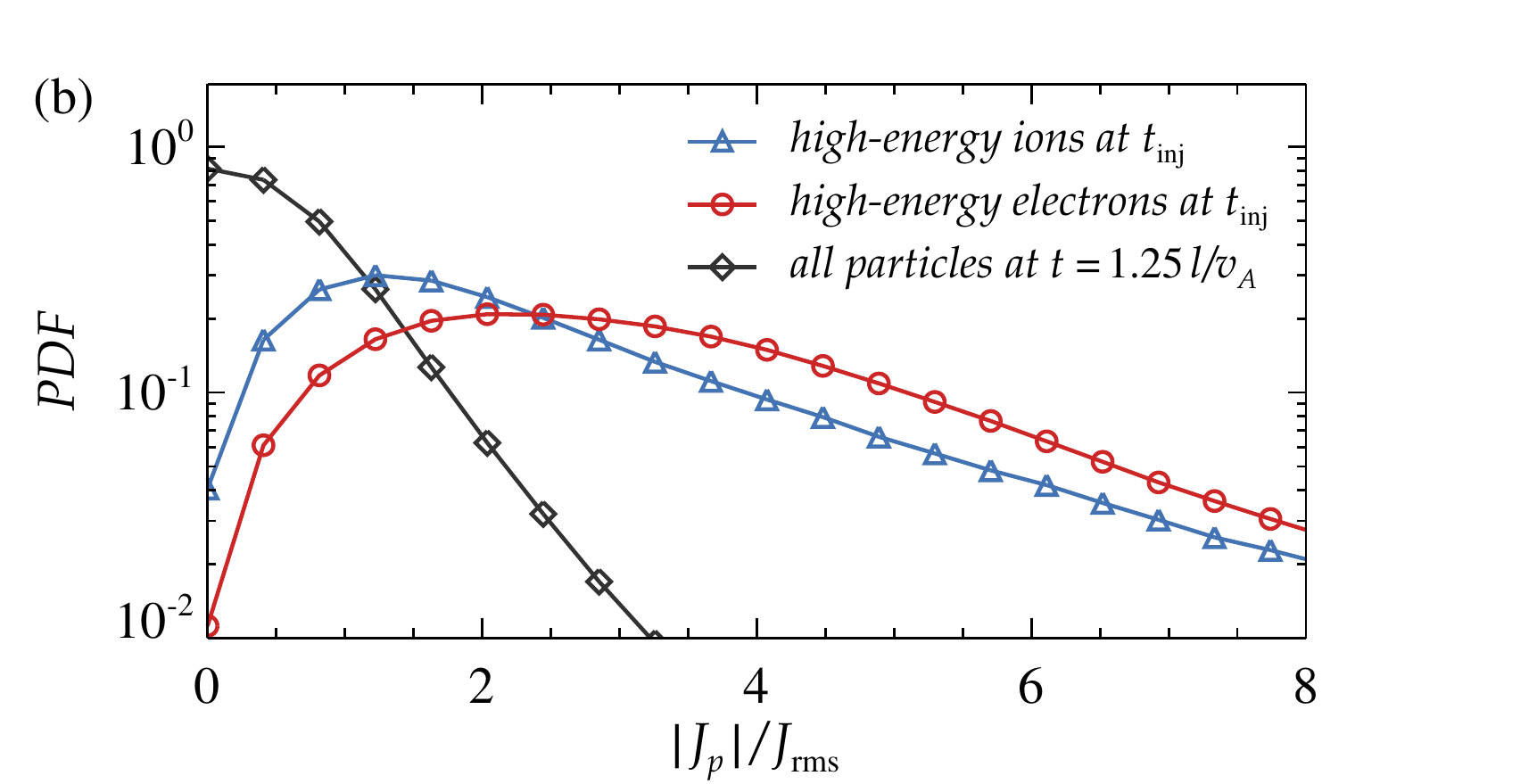}
\includegraphics[width=8.65cm]{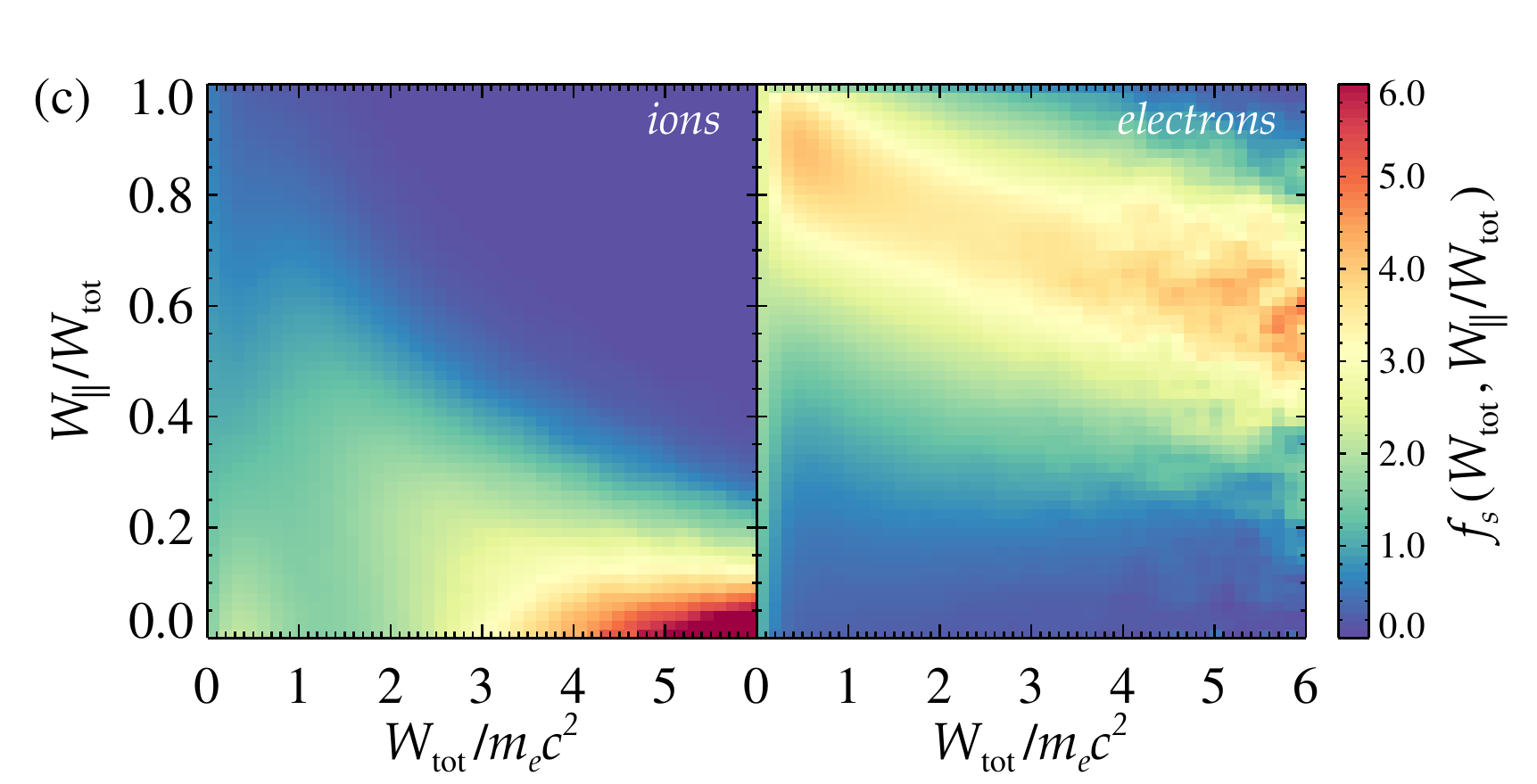}
\includegraphics[width=8.65cm]{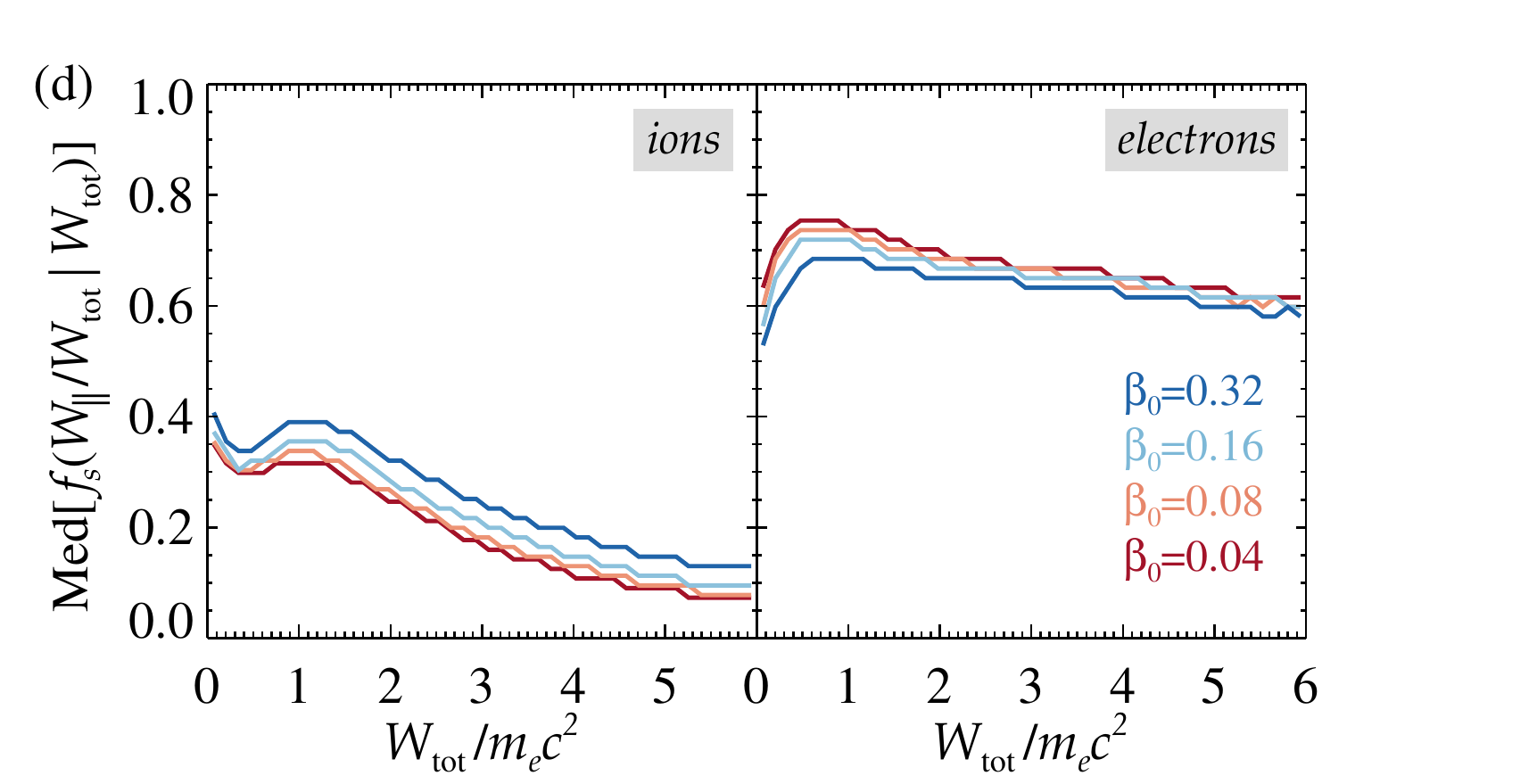}
\end{center}
\vspace{-0.2cm}
\caption{(a) Kinetic energy evolution for five representative ions (shades of blue) and five representative electrons (shades of red) that end up in the power-law range of the nonthermal tail. (b) PDFs of $|{J_{p}}|/{J_{{\rm{rms}}}}$ experienced by high-energy ions (blue triangles) and high-energy electrons (red circles) at their $t_{\rm{inj}}$ and by all our tracked particles at $t = 1.25 \, l/v_A$ (black diamonds). (c) Distribution of ions (left) and electrons (right) with respect to $W_{\rm{tot}}$ and $W_\parallel/W_{\rm{tot}}$, at the end of the $\beta_0=0.08$ simulation ($t = 8 \, l/v_A$). (d) Median of the conditional PDFs at given $W_{\rm{tot}}$ for ions (left) and electrons (right) at $t = 8 \, l/v_A$ from simulations with different $\beta_0$.}
\label{fig4}
\end{figure}

The significant energy dependence of the electron's pitch-angle distribution is clearly displayed in Fig.~\ref{fig3}(c), where we show $\langle \sin^2 \alpha \rangle$ as a function of $\varepsilon$ (normalized by $\varepsilon_*$) for different $\beta_0$ values. Contrary to ions, the electron $\langle \sin^2 \alpha \rangle$ deviates significantly from the expected mean for an isotropic distribution, $\langle \sin^2 \alpha \rangle = 2/3$ (compare with the black dashed line) and attains a minimum at $\varepsilon/\varepsilon_* \sim 2$ (as in pair plasma turbulence \citep{ComissoSironi19,Comisso20,ComissoSironi21}). The value of $\langle \sin^2 \alpha \rangle$ increases for higher energies and eventually $\langle \sin^2 \alpha \rangle \sim 2/3$ for $\varepsilon \gg \varepsilon_*$ as a consequence of pitch-angle scatterings off the turbulent fluctuations.

To understand the difference between the energization of ions and electrons, and its effect on pitch angle anisotropy, we tracked the trajectories of a random sample of $\sim{10^8}$ particles in each of the simulations. 
Figure~\ref{fig4}(a) shows the time evolution of the kinetic energy for 5 representative ions and 5 representative electrons that eventually populate the high-energy end of the nonthermal tail in the fiducial simulation. These particles experience a sudden ``injection phase,'' i.e., an episode of rapid energy gain that brings them to energies much higher than the initial thermal energy \citep{ComissoSironi18,ComissoSironi19}, followed by a more gradual energization phase akin to the stochastic Fermi acceleration mechanism \citep{fermi49}.

We find that the injection stage occurs at sites where the current density at the particle location, ${J_{p}}$, is significantly above the average value. This is illustrated in Fig.~\ref{fig4}(b), where we show the PDFs of $|{J_{p}}|/J_{{\rm{rms}}}$ at the injection time $t_{{\rm inj}}$ (for ions and electrons that eventually end up with high energies), and at a generic time. Here, we identified $t_{{\rm inj}}$ as the time when low-energy particles ($\varepsilon < 0.5 \, m_e c^2$) have a rate of increase of the particle kinetic energy (averaged over $\Delta{t} = 150 \, \omega_{pe}^{-1}$) that satisfies the empirical threshold $\Delta\varepsilon/\Delta{t} \ge 0.2 \, R_{{\rm rec}} v_A m_e c \, \omega_{pe}$, where $R_{{\rm rec}} = 0.1$ is the normalized collisionless reconnection rate \citep{ComissoJPP16,CassakJPP17}.  
For most particles, $t_{{\rm inj}}$ falls in the range $t v_A/l \sim 0.5-2$.  
One can see that the PDF of electrons at injection peaks at $|{J_{p}}|/{J_{{\rm{rms}}}} \sim 2.5$, while the PDF of ions at injection peaks at a lower value of $|{J_{p}}|/{J_{{\rm{rms}}}} \sim 1$. This should be contrasted with the PDF of the entire population of particles at a generic time, which peaks at $|{J_{p}}|/{J_{{\rm{rms}}}} \sim 0$. Therefore, particle injection is associated with locations of high current density (current sheets).

The injection phase of ions and electrons is governed by different energization mechanisms ($E_\parallel$ vs. $E_\bot$). In order to distinguish the role of the electric field parallel ($\parallel$) and perpendicular ($\bot$) to the local magnetic field, we compute ${W_\parallel}(t) = q \int_0^t {{{\bm{E}}_\parallel}(t') \cdot {\bm{v}}(t') \, dt'}$ and ${W_\bot}(t) = q \int_0^t {{{\bm{E}}_\bot}(t') \cdot {\bm{v}}(t') \, dt'}$ for all the tracked particles. This allows us to construct the distributions $f_s ({W_{\rm tot}}, {W_\parallel}/{W_{\rm tot}})$, where $W_{\rm tot} = W_\parallel + W_\bot$ is the work done by the total electric field. These distributions are shown in Fig.~\ref{fig4}(c) for ions (left) and electrons (right) from the fiducial simulation. 
Ions gain energy almost entirely via $E_\bot$. 
In contrast, the injection of  electrons is controlled by the $E_\parallel$ field associated with reconnecting current sheets, while $E_\bot$ energization from scatterings off the turbulent fluctuations becomes progressively more important as the electron energy increases.

The observed trend is robust across the different simulations. This is illustrated in Fig.~\ref{fig4}(d), where we show the median of $f_s ({W_\parallel}/{W_{\rm tot}} \, | {W_{\rm tot}})$ for various $\beta_0$ values. The contribution of $E_\bot$  consistently increases with energy in the range of the high-energy nonthermal tail,  for both ions and electrons. However, low-energy electrons need to be first accelerated via $E_\parallel$ up to the the characteristic energy $\varepsilon/\varepsilon_* \sim 2$ (Fig.~\ref{fig3}(c)) associated with magnetic reconnection. This is 
also the reason why the electron pitch-angle distribution develops a strong anisotropy, with $\min \langle \sin^2 \alpha \rangle$ attained at the same energy $\varepsilon/\varepsilon_* \sim 2$. For electrons, the initial energy gain via $E_\parallel$ increases with decreasing $\beta_0$, which is also consistent with the observed increase in pitch-angle anisotropy.

\section{Summary}
In summary, we have demonstrated with 3D fully kinetic simulations that magnetized plasma turbulence is a viable mechanism for ion and electron acceleration. Both species develop nonthermal power-law tails, but ions  attain harder nonthermal spectra and reach higher energies than  electrons. For both species, the power-law slope becomes harder when decreasing the plasma $\beta$, i.e., the ratio of  plasma pressure to  magnetic pressure.
We show that the energization of electrons is accompanied by a significant energy-dependent pitch-angle anisotropy, with most electrons moving parallel to the local magnetic field, while ions stay roughly isotropic. We demonstrate that particle injection from the thermal pool occurs in regions of high current density. Parallel electric fields associated with magnetic reconnection are responsible for the initial energy gain of electrons --- which also imprints into their pitch-angle anisotropy --- whereas perpendicular electric fields control the overall energization of ions.   The results of our first-principles simulations shed light on the origin of nonthermal particles in space and astrophysical systems.

$\,$

\clearpage

\acknowledgments
We acknowledge fruitful discussions with Daniel Gro\v{s}elj, Ramesh Narayan, Joonas N\"attil\"a, and Emanuele Sobacchi. We are particularly grateful to Nina McCurdy/NASA Ames for helping with the visualizations. This research acknowledges support from NASA 80NSSC18K1285, NSF PHY-1903412, DOE DE-SC0021254, and the Cottrell Fellowship Award RCSA 26932. The simulations were performed on Columbia University (Ginsburg), NASA-HEC (Pleiades), and NERSC (Cori) resources. 

$\,$

$\,$

\end{document}